\documentclass[% 
reprint,
amsmath,amssymb, 
aps,
prb
]{revtex4-2}
\usepackage{graphicx, color}% Include figure files
\usepackage{dcolumn}% Align table columns on decimal point
\usepackage{bm}% bold math
\usepackage[colorlinks,urlcolor=blue,linkcolor=blue,anchorcolor=blue,citecolor=blue]{hyperref}

\begin{document}
\preprint{APS/123-QED}
\title{Emergent $d$-wave altermagnetism in chlorine-adsorbed FeSe monolayer}
\author{Zi-Hao Ding$^{1}$}
\author{Ze-Feng Gao$^{1}$}
\author{Kai Liu$^{1}$}
\author{Peng-Jie Guo$^{1}$}
\email{guopengjie@ruc.edu.cn}
\author{Zhong-Yi Lu$^{1,2}$}
\email{zlu@ruc.edu.cn}
\affiliation{1. School of Physics and Key Laboratory of Quantum State Construction and Manipulation (Ministry of Education), Renmin University of China, Beijing 100872, China}
\affiliation{2. Hefei National Laboratory, Hefei 230088, China}
\date{\today}
\begin{abstract}
The recent emergence of altermagnetism has opened new frontiers in condensed matter physics, yet material platforms capable of hosting both intrinsic altermagnetic order and superconductivity remain exceedingly rare. Here, based on symmetry analysis and first-principles calculations, we propose a realistic route to engineer robust altermagnetism in monolayer FeSe, a prototypical iron-based superconductor. By designing a stoichiometric Fe$_2$Se$_2$Cl structure through single-side Cl adsorption and introducing gate-tunable hole doping, we achieve a highly stable altermagnetic ground state. Our calculations reveal a synergistic mechanism: hole doping firmly stabilizes the checkerboard magnetic order, while the asymmetric ligand environment intrinsically breaks the out-of-plane spatial inversion symmetry. Consequently, this interplay induces a giant altermagnetic spin splitting of up to 620 meV. Crucially, we demonstrate that this altermagnetic state and its giant spin splitting are highly resilient, persisting even in a 10-layer slab model that accurately simulates the bulk limit. By introducing altermagnetism into the well-established FeSe-based superconducting family, our findings identify Fe$_2$Se$_2$Cl as a promising platform for spintronic applications and motivate future studies of the possible interplay between altermagnetism and superconductivity.
\end{abstract}

\maketitle

%\tableofcontents
\section{Introduction}
%\textit{Introduction.}
Altermagnetism has recently emerged as a fundamentally distinct magnetic phase, uniquely combining the fully compensated macroscopic magnetization of antiferromagnets in real space with the momentum-space spin polarization characteristic of ferromagnets \cite{PRX-1,PRX-2}. Driven by either crystal potentials or spontaneous correlations, altermagnets exhibit unconventional anisotropic (e.g., $d$-, $g$-, or $i$-wave) alternating spin polarization without relying on relativistic spin-orbit coupling \cite{orbital}. This distinctive electronic structure fundamentally breaks macroscopic time-reversal symmetry, giving rise to a host of novel phenomena, including spin currents \cite{SST-PRL2021,JunweiLiu,SST-NE2022,SST-PRL2022,SST-PRL2022-2}, a diverse family of anomalous Hall effects \cite{sciadv.aaz8809,vsmejkal2022anomalous,npj,CTT,CVHE-tan,Type-II-QSHI,TAN2026,excitonicquantumanomaloushall}, and ferroelectricity \cite{Ferroelectric-1,Ferroelectric-2,Ferroelectric-3,Ferroelectric-4}. Furthermore, altermagnetism can induce novel density wave phases \cite{KV2Se2O,CsCr3Sb5,magnetic-field-tunable,ding2025}, further enriching the landscape of correlated phenomena. %Consequently, altermagnetism not only provides a highly promising platform for next-generation spintronics but also opens new avenues for exploring its rich interplay with other states of matter. 
In particular, its interplay with superconductivity and altermagnetic fluctuations is being actively investigated—both as an intrinsic phenomenon in bulk systems where pairing and altermagnetic order coexist, and in heterostructures where it is induced via the proximity effect \cite{mazin2025notes,superconductivity-1,superconductivity-2,superconductivity-3,superconductivity-4,superconductivity-5,superconductivity-6,superconductivity-7}. In a more fundamental sense, these materials can be viewed as the magnetic counterpart to unconventional $d$-wave superconductors, deepening the profound physical analogy between magnetism and superconductivity.

Indeed, a compelling connection between altermagnetism and high-temperature superconductivity is rapidly emerging in canonical material families. For example, La$_2$CuO$_4$, the parent compound of cuprate superconductors, has been identified as an altermagnet \cite{PRX-1}. Recent first-principles studies have revealed that high-pressure vacancy-ordered iron selenides (A$_2$Fe$_4$Se$_5$ phase) undergo a magnetic phase transition from a block-spin antiferromagnetic phase to an intrinsic Néel altermagnetic phase \cite{A2Fe4Se5}. However, material candidates for altermagnetism-related superconductivity (SC) are currently still lacking and remain elusive. In this context, FeSe, the structurally simplest iron-based superconductor, emerges as an exceptionally promising platform. It hosts remarkably rich physics, where its unique nonmagnetic nature provides a challenging testing ground for conventional spin-fluctuation pairing mechanisms. Furthermore, the superconducting transition temperature ($T_c$) of FeSe exhibits striking tunability. The $T_c$ of FeSe, while below 8 K at ambient pressure \cite{PNAS}, reaches 36.7 K under an applied pressure of 8.9 GPa \cite{36.7}. Owing to its weak interlayer van der Waals bonding, FeSe is easily processed via mechanical exfoliation or chemical intercalation. Moreover, carrier density in FeSe flakes can be drastically tuned through ionic liquid gating, which further enhances the superconductivity up to 43 K \cite{Flakes}. Monolayer FeSe grown on SrTiO$_3$ substrates by molecular beam epitaxy (MBE) has achieved high-temperature superconductivity exceeding 65 K \cite{2012,2013,2013interface}, opening exciting new avenues for exploring two-dimensional interfacial superconductivity enhancement. Theoretical predictions suggest that applying an out-of-plane electric field to monolayer FeSe with a checkerboard magnetic order can induce an altermagnetic state accompanied by a quantized spin Hall conductivity \cite{induced}. However, in monolayer FeSe, the checkerboard order is merely one of several competing magnetic phases, and the primary role of the applied electric field is simply to render the top and bottom ligand layers chemically distinct. Such an approach relies heavily on extrinsic perturbations and specific metastable states. Consequently, a natural question arises: can robust altermagnetism be induced in monolayer FeSe?

In this work, we propose a practical route to realize a stable altermagnetic phase in monolayer FeSe through single-side Cl adsorption—forming a stoichiometric Fe$_2$Se$_2$Cl structure—combined with hole doping via ionic gating. Our systematic investigations reveal that hole doping effectively stabilizes the checkerboard magnetic order as the robust ground state. Concurrently, the asymmetric Cl adsorption permanently distinguishes the top and bottom ligand environments, intrinsically breaking the out-of-plane spatial inversion symmetry. This synergistic interplay naturally induces a robust altermagnetic state, yielding a giant altermagnetic spin splitting of up to 620 meV. Crucially, this altermagnetism and its associated spin splitting remain highly robust even when the model is extended to a 10-layer slab to accurately simulate the realistic bulk limit. More fundamentally, because FeSe is a prototypical iron-based superconductor, integrating altermagnetism into this specific material family paves the way for exploring the elusive coexistence of altermagnetic order and superconductivity. This ultimately establishes a highly viable material candidate for future topological quantum computing and spintronic applications.

\begin{figure}[htbp]
	\centering
	\includegraphics[width=8.5cm]{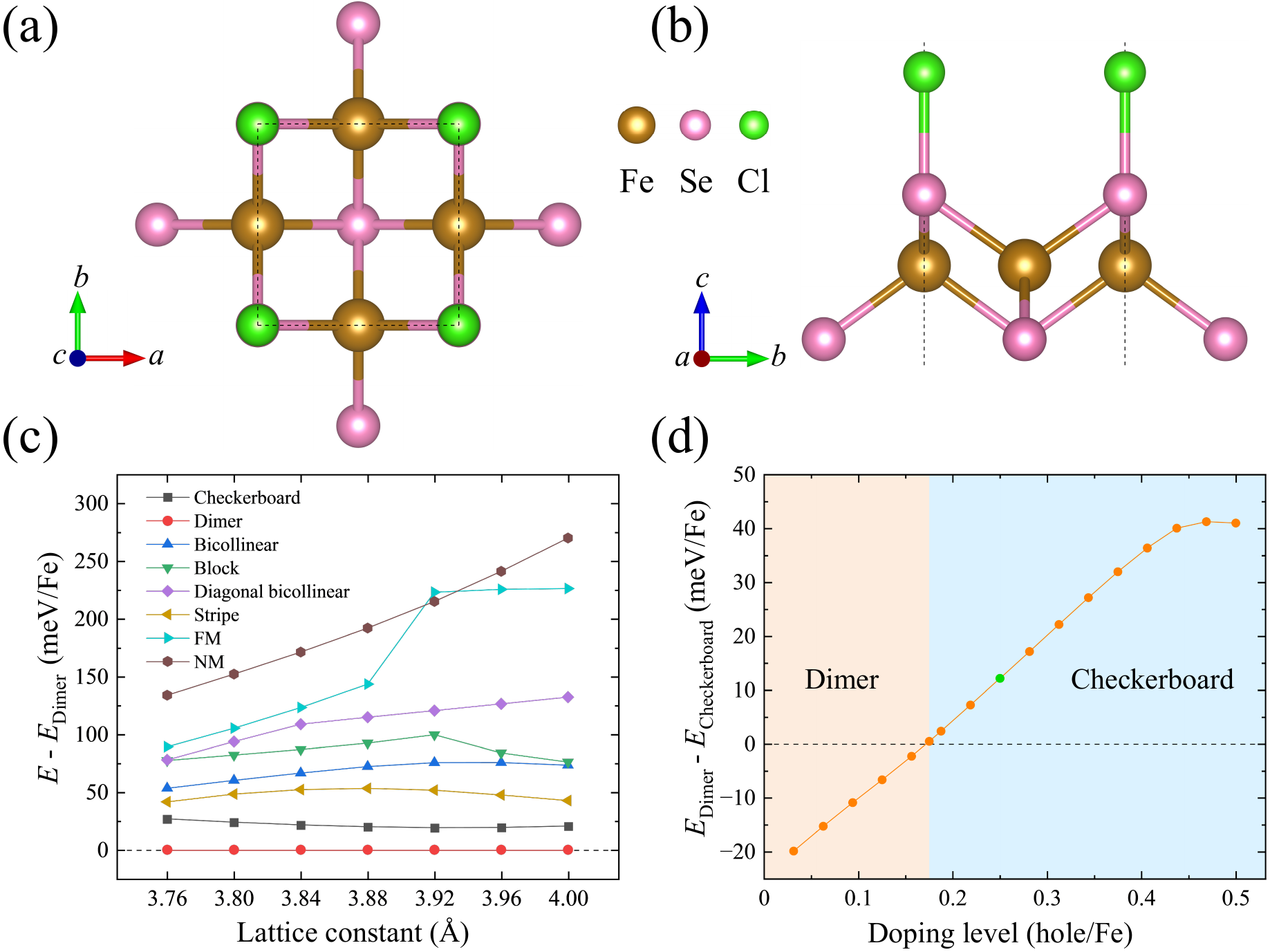}
	\caption{(a) Top and (b) side views of the monolayer Fe$_2$Se$_2$Cl crystal structure. (c) Relative energies of various magnetic configurations as a function of the lattice constant, calculated with respect to the dimer order. (d) Energy difference between the dimer and checkerboard magnetic orders, defined as $\Delta E = E_{\text{dimer}} - E_{\text{checkerboard}}$, as a function of the hole doping level.}
	\label{fig1}
\end{figure}

\section{Results and analysis}
%\textit{Results and analysis.}
The monolayer Fe$_2$Se$_2$Cl crystallizes in a tetragonal lattice structure belonging to the space group $P4mm$ (No.~99) with $C_{4v}$ point group symmetry. As illustrated in Figs.~\ref{fig1}(a) and \ref{fig1}(b), the primitive cell is formed by an FeSe monolayer with single-sided Cl adsorption directly above the top Se atom. Consequently, the inversion symmetry $I$ is broken, while the $C_{4z}$ rotational symmetry is preserved. The fully optimized lattice constant is calculated to be 3.796~\AA, which is in good agreement with previous experimental reports for single unit-cell FeSe grown on SrTiO$_3$ (3.82~\AA) \cite{3.82} and bulk FeSe (3.765~\AA) \cite{PNAS}. The overall structure of the FeSe monolayer is well preserved within Fe$_2$Se$_2$Cl, suggesting the potential for similar superconducting behavior.

\begin{figure}[htbp]
	\centering
	\includegraphics[width=8.5cm]{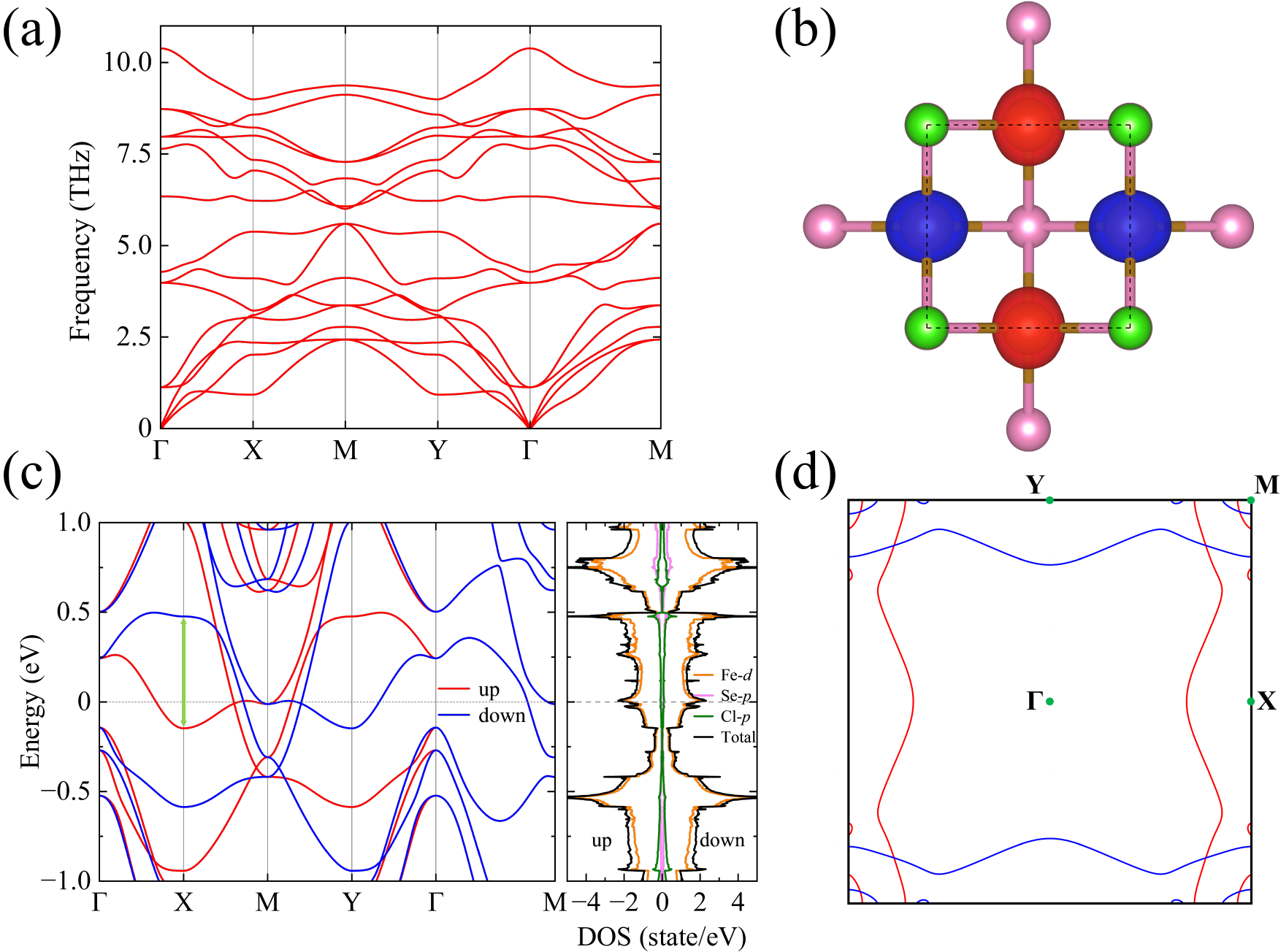}
	\caption{(a) Phonon spectrum of monolayer Fe$_2$Se$_2$Cl along high-symmetry paths, calculated for the checkerboard altermagnetic order. (b) Distribution of the spin density, where red and blue isosurfaces represent the spin-up and spin-down densities, respectively. (c) Electronic band structure and the corresponding density of states (DOS) calculated without spin-orbit coupling (SOC). The large altermagnetic spin splitting of $\sim$620~meV is highlighted by the light green arrow. (d) Spin-resolved Fermi surface.}
	\label{fig2}
\end{figure}

To investigate the stability of the monolayer Fe$_2$Se$_2$Cl, we first examine six different adsorption sites, as shown in Fig.~S1 of the Supplemental Material~\cite{SM}. Our calculations identify Site~1, where the Cl atom is positioned directly above the top Se atom with a bond length of 2.4~\AA, as the most energetically favorable adsorption site. Subsequently, we investigate four different Cl coverages (0.125, 0.25, 0.5, and 1 ML) using a $2\sqrt{2} \times 2\sqrt{2} \times 1$ supercell. Here, a coverage of 1 ML is defined as the atomic density of the top Se layer. As illustrated in Fig.~S2, we consider three representative atomic configurations for both the 0.25 and 0.5 ML cases. After structural optimization,
we calculate the adsorption energies at various coverages. The negative values indicate that the adsorption is an exothermic process. Importantly, the adsorption energy becomes progressively more negative with increasing coverage, suggesting that the system reaches its most stable state when all available positions are fully occupied (Table~S1). Furthermore, our phonon calculations [Fig.~\ref{fig2}(a)] and molecular dynamics simulations (Fig.~S3) demonstrate the dynamical and thermal stabilities of the structure. At 300~K, no significant distortion, disorder, clustering, or structural phase transitions are observed. We therefore conclude that a stable, stoichiometric 2D material, Fe$_2$Se$_2$Cl, is formed via single-sided Cl adsorption on the monolayer FeSe.

To investigate the magnetic properties of the monolayer Fe$_2$Se$_2$Cl, we consider eight different magnetic configurations based on a $2\sqrt{2} \times 2\sqrt{2} \times 1$ supercell, as illustrated in Fig.~S4. Initially, we perform total-energy calculations as a function of the lattice constant over the range from 3.76~\AA\ to 4.00~\AA. This range encompasses the lattice constants of both bulk FeSe (3.765~\AA) and the SrTiO$_3$ (STO) substrate (3.905~\AA) \cite{3.905}. As depicted in Fig.~\ref{fig1}(c), within the lattice constant range of 3.76~\AA\ to 4.00~\AA, the dimer order consistently exhibits the lowest energy, followed by the checkerboard order. At the fully optimized lattice constant of 3.796~\AA, the total energy of the dimer order is calculated to be 24.186 meV/Fe lower than that of the checkerboard order. However, from the perspective of spin group symmetry, the dimer order in the monolayer Fe$_2$Se$_2$Cl does not classify as an altermagnet, because the two opposite spin sublattices are related by a lattice translation. Given that the system intrinsically breaks inversion symmetry, it is only within the checkerboard order that the two spin sublattices can be connected by a crystalline rotation or mirror symmetry, thereby making the material a textbook altermagnet.

As an effective approach for tuning magnetic properties, carrier doping has been established as a pivotal route to stabilize metallic altermagnetism in correlated materials with competing magnetic interactions \cite{mazin2021prediction,Co-dopedFeSb2}. This is particularly viable given that systematic modulation of carrier density via liquid gating has been successfully employed to drive the evolution of superconductivity in FeSe thin films \cite{Flakes}. We calculate the magnetic ground states across various hole doping levels. The dimer and checkerboard configurations consistently remain the two lowest-energy orders across all considered doping levels. As depicted in Fig.~\ref{fig1}(d), we find that with progressive hole doping, the magnetic ground state undergoes a transition from the dimer order to the checkerboard order, which intrinsically hosts altermagnetism.

In the following, we choose the specific doping level of 0.25 hole/Fe indicated in green in Fig.~\ref{fig1}(d) as a representative example to elucidate the altermagnetic properties of Fe$_2$Se$_2$Cl. At this doping level, the total energy of the checkerboard altermagnetic order is 12.19 meV/Fe lower than that of the dimer order. The anisotropic electric crystal potential introduced by the Cl atoms gives rise to anisotropic exchange interactions, resulting in a highly anisotropic spin density within the system, as shown in Fig.~\ref{fig2}(b). In addition, the two Fe sublattices, characterized by antiparallel magnetic moments, are related by spin group symmetry operations $\{C_2 || C_{4z}\}$, $\{C_2 || M_{d}\}$, and $\{C_2 || M_{d\perp}\}$. Here, the $M_d$ and $M_{d\perp}$ denote mirror reflections with respect to the diagonal axes $\hat{d} = \hat{x} + \hat{y}$ and $\hat{d}_\perp = \hat{x} - \hat{y}$, respectively. These symmetry operations satisfy the fundamental requirements for altermagnetism. Consequently, this real-space anisotropy is anticipated to induce a large momentum-dependent spin splitting. Our electronic band structure calculations corroborate this hypothesis. As depicted in Fig.~\ref{fig2}(c), in the absence of spin-orbit coupling (SOC), the electronic structure is metallic, with partially filled Fe $d$ orbitals dominating the states around the Fermi level. The density of states for the spin-up and spin-down channels are perfectly symmetric, indicating a fully compensated macroscopic magnetization. $\text{Fe}_2\text{Se}_2\text{Cl}$ indeed exhibits a distinct momentum-dependent spin splitting, confirming its altermagnetic order. Near the Fermi level, a significant splitting of up to nearly 620 meV is observed between the spin-up and spin-down channels along the $\Gamma$-X(Y)-M paths. Crucially, the spin polarization exactly reverses when rotating from the $\Gamma$-X-M to the $\Gamma$-Y-M paths. The calculated spin-resolved Fermi surface, shown in Fig.~\ref{fig2}(d), exhibits a planar $d$-wave symmetry.

\begin{figure}[htbp]
	\centering
	\includegraphics[width=8.5cm]{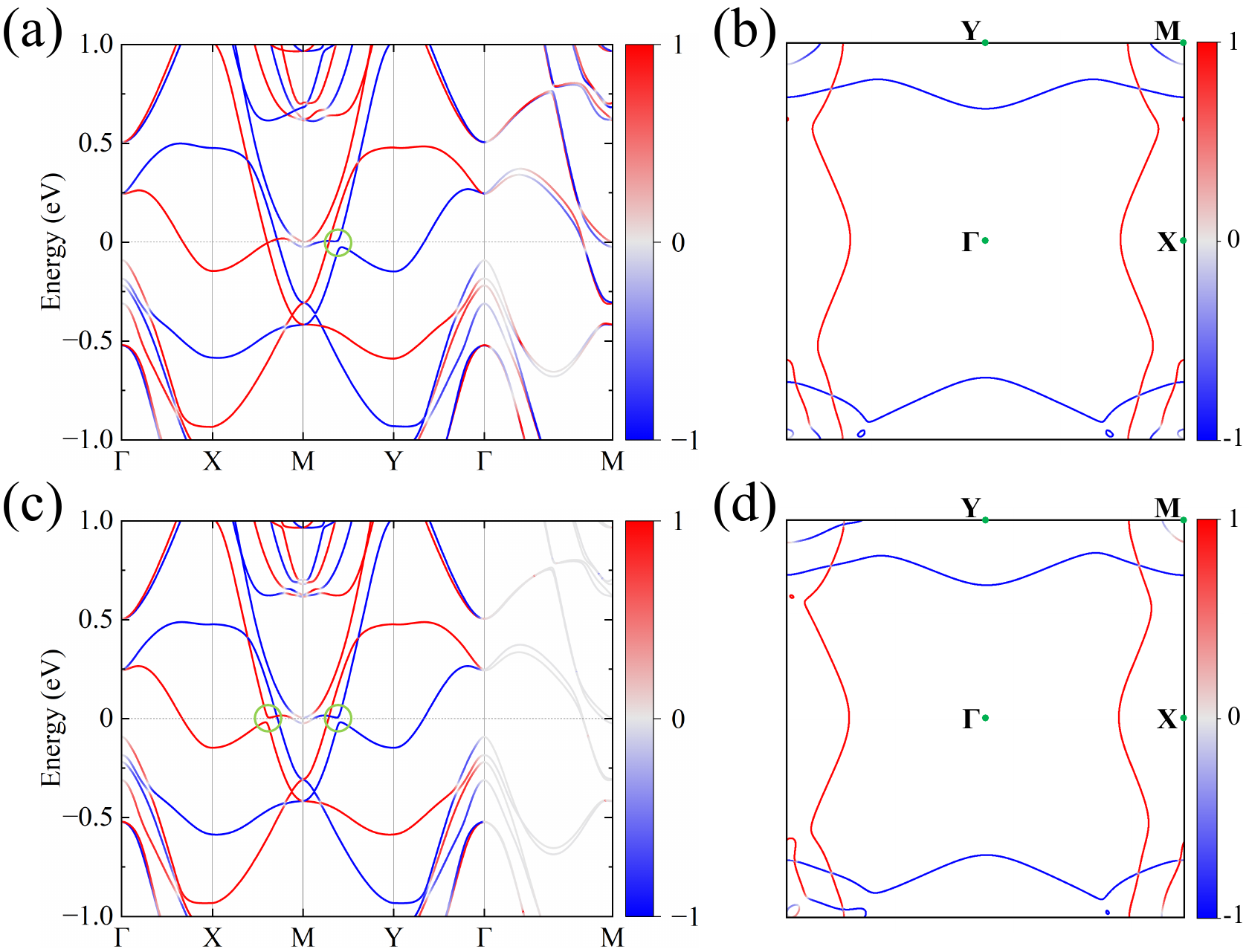}
	\caption{(a) Spin-projected energy bands and (b) corresponding Fermi surface calculated with spin-orbit coupling (SOC) for the N\'eel vector pointing along the [100] direction. (c) Energy bands and (d) Fermi surface calculated with N\'eel vector oriented along the [110] direction. The color scale in all panels represents the magnitude of the spin projection. The anti-crossings near the Fermi level are highlighted by light green circles.}
	\label{fig3}
\end{figure}

The inclusion of SOC in our calculations reveals that the altermagnetic order in Fe$_2$Se$_2$Cl exhibits an easy-plane magnetic anisotropy. We find that the ground-state energy is identical when the N\'{e}el vector points along the [100], [110], and [010] directions.
We calculate the electronic band structures and Fermi surfaces with spin projections along the [100] and [110] directions, respectively. As shown in Figs.~\ref{fig3}(a) and \ref{fig3}(c), the introduction of SOC has a minimal impact on the overall band structure. However, the most striking feature is that anti-crossings appear at distinct points near the Fermi level. Specifically, for the [100] spin orientation, the Weyl point along the X-M path remains gapless, whereas a gap opens at the Weyl point along the M-Y path. In contrast, when the spins are aligned along the [110] direction, both Weyl points become gapped. This is due to the fact that different spin quantization axes lead to distinct magnetic space groups: $P m^\prime m 2^\prime$ for the [100] direction and $C m^\prime m 2^\prime$ for the [110] direction. Along the $[100]$ direction, the system preserves the symmetry operations $\{E, M_x\} + \mathcal{T}\{C_{2z}, M_y\}$, where $\mathcal{T}$ is the time-reversal operator. Similarly, the symmetry operations along the $[110]$ direction are given by $\{E, M_d\} + \mathcal{T}\{C_{2z}, M_{d\perp}\}$. The Fermi surfaces and their corresponding spin projections along the [100] and [110] directions, as depicted in Figs.~\ref{fig3}(b) and \ref{fig3}(d), strictly obey the aforementioned symmetry constraints.

\begin{figure}[htbp]
	\centering
	\includegraphics[width=8.5cm]{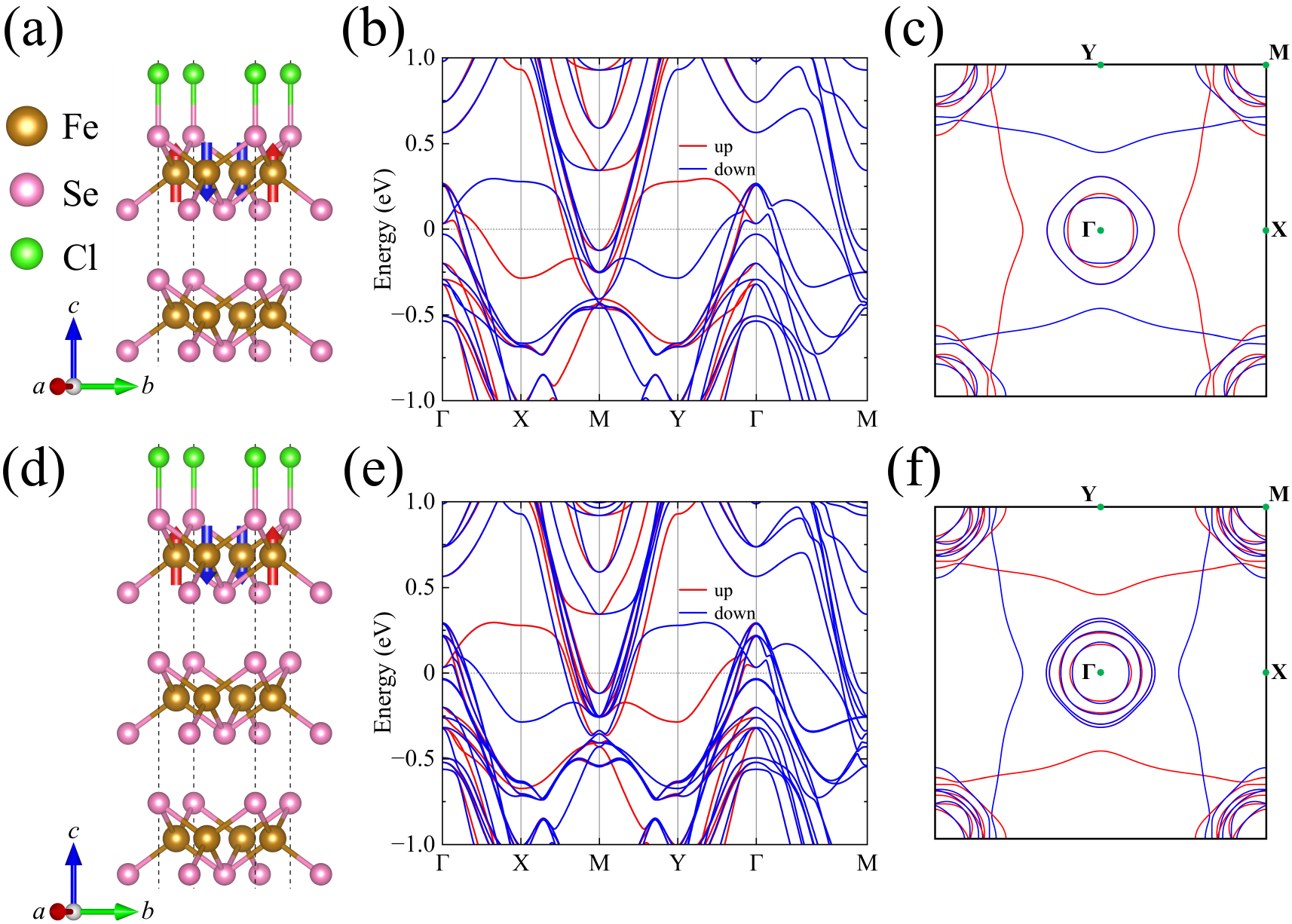}
	\caption{(a) Electronic band structure and (b) spin-resolved Fermi surface calculated without spin-orbit coupling (SOC) for the bilayer system (one Fe$_2$Se$_2$Cl layer on one FeSe layer). (c) Electronic band structure and (d) spin-resolved Fermi surface for the trilayer system (one Fe$_2$Se$_2$Cl layer on two FeSe layers).}
	\label{fig4}
\end{figure}

Motivated by the experimental absence of static long-range magnetic order in bulk and thin-film FeSe, we further investigate the influence of the nonmagnetic underlying FeSe layers on the altermagnetism in Fe$_2$Se$_2$Cl. To circumvent the computational limitation of applying layer-specific hole doping in standard plane-wave frameworks, we explicitly initialize the top Fe$_2$Se$_2$Cl layer with a checkerboard altermagnetic order, while setting the initial magnetic moments of all underlying FeSe layers to zero. Upon structural relaxation with interlayer van der Waals (vdW) corrections for the bilayer and trilayer systems, the underlying layers reliably converge to a nonmagnetic state, whereas the top-layer Fe atoms retain their local magnetic moments. As shown in Fig.~\ref{fig4}, the characteristic altermagnetic spin splitting remains remarkably stable in this configuration. Crucially, even when the model is extended to a 10-layer slab to accurately simulate the true bulk limit [Fig.~S5], this altermagnetic spin splitting persists robustly. These results unambiguously confirm that the altermagnetic state in this system is an intrinsic surface-driven phenomenon. It is highly robust against interlayer hybridization and, most importantly, completely decoupled from the bulk nonmagnetic state. 
%This strongly suggests that the predicted altermagnetic spin splitting should be experimentally observable, even in the absence of bulk long-range magnetic order.

\section{Discussion and conclusion}
%\textit{Discussion and conclusion.}
From an experimental perspective, the realization of Fe$_2$Se$_2$Cl is highly feasible. The surface functionalization and halogenation of 2D materials have been widely demonstrated via chemical vapor deposition or controlled surface treatment \cite{covalent,small}. Furthermore, hole doping in iron-based superconductors is a mature technique, frequently achieved through non-destructive ionic liquid gating \cite{Flakes}. The doping concentration of 0.25 hole per Fe atom (equivalent to $3.47 \times 10^{14}~\text{ hole/cm}^2$) is experimentally viable, particularly in light of established reports demonstrating that doping concentrations up to $3.5 \times 10^{15}~\text{e/cm}^2$ can be realized through electrochemical gating techniques \cite{3.5}. 
Indeed, recent theoretical studies have extensively highlighted the tremendous potential of surface decoration in tailoring the physical properties of FeSe-based systems. For instance, it has been demonstrated that surface fluorine (F) adsorption on a superconducting FeSe monolayer can effectively control its magnetic anisotropy and induce ferroelasticity \cite{FeSeF}. Beyond halogenation, decorating iron-based superconductors with lithium (Li) has been predicted to realize high-temperature quantum anomalous Hall insulators \cite{LiFeSe}, while the introduction of metal atoms (Tl, In, Ga) to form 2D MFeSe systems can yield tunable Weyl half-semimetals \cite{MFeSe}. These compelling theoretical precedents strongly validate the paradigm of reconstructing the electronic and magnetic structures of FeSe via surface functionalization.

The potential coexistence of altermagnetism and proximity-induced superconductivity in Fe$_2$Se$_2$Cl opens a fascinating frontier. In conventional superconductors, magnetic order typically acts as a pair-breaker. However, the unique $d$-wave symmetry of the altermagnetic spin polarization in momentum space can lead to novel superconducting states. Specifically, the lifted spin degeneracy in the normal state of Fe$_2$Se$_2$Cl naturally biases the system toward equal-spin triplet pairing ($\Delta_{\uparrow\uparrow}$ or $\Delta_{\downarrow\downarrow}$). This potential for intrinsic triplet superconductivity is a key requirement for realizing Majorana zero modes, positioning this material as a promising candidate for topological quantum computation.

During the preparation of this manuscript, we became aware of two contemporaneous studies. One proposed that single-sided hydrogenation, oxidation, and fluorination can transform spin-degenerate antiferromagnetic FeSe into a $d$-wave altermagnet with pronounced momentum-dependent spin splitting~\cite{Cui2026Surface}. The other reported the coexistence of $d$-wave altermagnetism and nontrivial topology in Janus FeSe$X$ ($X=\mathrm{S,Te}$) monolayers~\cite{GonzalezGarcia2026Janus}. Although these results are consistent with the symmetry analysis presented here, observing altermagnetic spin splitting in a prescribed checkerboard configuration does not by itself establish that this configuration is the true magnetic ground state. The former study does not report a systematic total-energy comparison among competing magnetic orders, whereas the latter compares only the ferromagnetic, checkerboard altermagnetic, and nonmagnetic states within the primitive cell. Consequently, magnetic states with longer periodicities cannot be excluded.

By explicitly comparing the relevant magnetic configurations in a $2\sqrt{2} \times 2\sqrt{2} \times 1$ supercell, we find that the ideal undoped Fe$_2$Se$_2$Cl monolayer instead adopts a dimer-ordered ground
state, which lies 24.186 meV/Fe below the checkerboard order. Hole doping reverses this energy hierarchy at approximately 0.175 hole/Fe, and at 0.25 hole/Fe the checkerboard order becomes lower than the dimer order by 12.19 meV/Fe. In realistic samples, epitaxial strain, substrate hybridization, interfacial charge transfer, defects, nonuniform adsorption, and finite-temperature effects may alter the relative
stability of the competing magnetic orders and thereby stabilize the checkerboard altermagnetic order even without intentional carrier doping. Our results therefore do not exclude the realization of
altermagnetism in nominally undoped experimental samples. Rather, they identify hole doping as a controllable and reliable route to energetically stabilize the checkerboard altermagnetic ground state,
whereas asymmetric surface functionalization creates the symmetry environment required for altermagnetic spin splitting.

In summary, our first-principles study identifies the hole-doped Fe$_2$Se$_2$Cl system as a stoichiometric, bulk-compatible 2D altermagnetic platform. We thereby provide a concrete material platform that integrates altermagnetism with potential high-temperature superconductivity. Ultimately, this work not only expands the library of 2D altermagnetic materials but also provides a clear roadmap for designing stray-field-free spintronic devices and exploring exotic quantum phases in iron-based heterostructures.

\begin{acknowledgments}
%\textit{Acknowledgments.}
This work was financially supported by the National Natural Science Foundation of China (Grant No.12434009, No.62476278 and No.12174443), the National Key R$\&$D Program of China (Grant No. 2024YFA1408601),the Fundamental Research Funds for the Central Universities, and the Research Funds of Renmin University of China (Grant No. 24XNKJ15). Computational resources have been provided by the Physical Laboratory of High Performance Computing at Renmin University of China.
\end{acknowledgments}

%\appendix
%\begin{appendices}

%\section{Some details for calculating the N\'eel temperatures\label{appA:appendix A}}

%\end{appendices}

\nocite{*}

\bibliography{Reference}% Produces the bibliography via BibTeX.

\end{document}